\documentclass{article}

\usepackage{graphicx}

\usepackage{graphicx}% Include figure files
\usepackage{dcolumn}% Align table columns on decimal point
\usepackage{bm}% bold math
\usepackage{amsfonts}
\usepackage{amsmath}

\usepackage{float}
\makeatletter

\newcommand{\Rmnum}[1]{\expandafter\@slowromancap\romannumeral #1@}
\makeatother
\begin{document}

\title{Entropy bound for the photon gas in noncommutative spacetime}
\author{K. Nozari$^{1}$\thanks{email: knozari@umz.ac.ir},
\hspace{.2cm} M. A. Gorji$^{1}$\thanks{email:
m.gorji@stu.umz.ac.ir}, \hspace{.2cm} A. Damavandi
Kamali$^2$\thanks{email: adkamali@srbiau.ac.ir} \hspace{.2cm} and
\hspace{.2cm} B. Vakili$^3$\thanks{email: b-vakili@iauc.ac.ir}
\vspace{.3cm}\\ $^1${\small {\it Department of Physics, Faculty of
Basic Sciences, University of Mazandaran,}}\\{\small {\it P. O. Box
47416-95447, Babolsar, Iran.}}\\$^2${\small {\it Department of Physics,
Science and Research Branch, Islamic Azad University, Tehran, Iran}}\\
$^3${\small {\it Department of Physics, Central Tehran Branch,
Islamic Azad University, Tehran, Iran}}}\maketitle

\begin{abstract}
Motivated by the doubly special relativity theories and
noncommutative spacetime structures, thermodynamical properties of
the photon gas in a phase space with compact spatial momentum space
is studied. At the high temperature limit, the upper bounds for the
internal energy and entropy are obtained which are determined by the
size of the compact spatial momentum space. The maximum internal
energy turns out to be of the order of the Planck energy and the
entropy bound is then determined by the factor $\big(V/l_{_{\rm
Pl}}^3\big)$ through the relevant identification of the size of the
momentum space with Planck scale. The entropy bound is very similar
to the case of Bekenstein-Hawking entropy of black holes and
suggests that thermodynamics of black holes may be deduced from a
saturated state in the framework of a full quantum gravitational
statistical mechanics.\\
\begin{description}
\item[PACS numbers]
04.60.Bc, 05.20.-y, 65.40.gd
\item[Key Words]
Quantum Gravity Phenomenology, Noncommutative spacetime, Photon Gas,
Entropy
\end{description}
\end{abstract}
\section{Introduction}
Existence of a minimum length scale, below which no other length
could be observed, is suggested by all quantum gravity candidates
such as string theory and loop quantum gravity \cite{String,LQG}. It
is then widely believed that a non-gravitational theory which
contains a universal minimal length scale will be arisen at the flat
limit of quantum gravity. Although the standard relativistic quantum
mechanics does not take into account any minimal length scale, the
flat limit of quantum gravity will be reduced to this standard
theory in the continuum limit (if it is exist) in the light of
correspondence principle. In this respect, the deformation of
standard relativistic quantum mechanics seems to be necessary in
order to take into account a quantum gravity length scale. The first
attempt in this direction goes back to the seminal work of Snyder
who formulated a discrete Lorentz-invariant spacetime in which the
spacetime coordinates appear to be non-commuting operators
\cite{Snyder}. On the other hand, it is well-known that the
transition from non-relativistic to relativistic and also classical
to quantum mechanics are obtained by stabilizing deformations of two
unstable algebraic structures, say, Lie algebra of Galilean group
and Poisson algebra, to two robust stable Lorentz and Heisenberg
algebras. However, when Lorentz and Heisenberg stable algebras are
put together, the resultant algebra for the relativistic quantum
mechanics turns out to be unstable. Thus, it is natural to explore a
stable algebra for the relativistic quantum mechanics
\cite{Stable-Th}. Evidently, the spacetime coordinates also become
non-commuting operators for the resultant stable algebra of the
relativistic quantum mechanics and a parameter with dimension of
length naturally arises in this setup \cite{Mendes}. One could then
consider this noncommutative stable algebra to be a candidate for
the flat limit of quantum gravity by the relevant identification of
the associated deformation parameter with universal minimal length.
The non-commutativity of spacetime coordinates is also suggested by
string theory at fundamental level \cite{S-NC}. The direct
consequence of the non-commutating spacetime coordinates is the
modification of the Heisenberg uncertainty principle (see for
instance \cite{Mendes-UR}). In this respect, one could also study
the effects of a universal minimal length to the classical (special
relativity) and non-relativistic (quantum mechanics) limit of
relativistic quantum mechanics. For instance, inspired by string
theory, generalized uncertainty relations are suggested \cite{GUP0}
which support the existence of a minimal length through a nonzero
uncertainty in position measurement \cite{GUP}. This deformed
quantum mechanics is very similar to the one that is inspired by
non-relativistic subalgebra of Snyder noncommutative algebra
\cite{Mignemi}. Furthermore, from the fact that a minimal length in
one inertial frame may be different in another observer's frame
through the Lorentz-Fitzgerald contraction in special relativity,
the doubly special relativity (DSR) theories are investigated in
order to take into account an observer-independent minimum length
(or maximal energy) as well as velocity of light \cite{DSR}. The
Lorentz invariance then may be considered only as an approximate
symmetry which will be broken at the Planck scale. However, it could
be possible to formulate a Lorentz-invariant DSR theory through a
nonlinear action of Lorentz group on four-momentum space
\cite{DSR-LI}. Recently, it is found that the DSR theories can be
realized from maximally symmetric curved four-momentum spaces and
the observer-independent scale is determined by the constant
curvature of the corresponding four-momentum space in this setup
\cite{Curve-M}. The duality of curved momentum space and
noncommutative spacetime is also shown in quantum geometry setup
\cite{Majid}. Generally, deformed noncommutative spacetimes could be
realized from deformed $\kappa$-Poincar\'{e} algebra on the
associated $\kappa$-Minkowski noncommutative spacetime
\cite{Poincare}. For instance, the Lorentz-violating algebra that is
investigated in Ref. \cite{DSR} and a Lorentz-invariant version that
is proposed later in Ref. \cite{DSR-LI}, could be realized from the
different bases of $\kappa$-Poincar\'{e} algebra in unified picture
\cite{Glikman}. Also, it is shown that the Snyder algebra
\cite{Snyder} and the stable algebra of relativistic quantum
mechanics \cite{Mendes} can be obtained from a ten-dimensional phase
space, with curved geometry for four-momentum space, through the
symplectic reduction process \cite{Girelli}. In these respects,
curved four-momentum spaces found their robust role in the relative
locality principle framework \cite{R-L}.

The direct consequence of the curved four-momentum space or
noncommutative spacetime is the deformation to the dispersion
relation and the associated density of states
\cite{DOS,DSR-THR,DOS-Ads}. Since density of states determines the
number of microstates in statistical mechanics, the thermodynamical
properties of the physical systems in these setups would be
significantly different from the standard ones. Indeed, existence of
a minimal length and maximal momentum (or maximal energy), as an
extra information for the system under consideration, change the
standard uniform probability distribution of microstates at high
energy regime. In recent years, the effects of minimal length and
maximal momentum on thermodynamical systems are studied in many
contexts. For instance, thermodynamical properties of some physical
systems in noncommutative spaces are studied in Ref. \cite{NC-THR}.
For the case of generalized uncertainty principle, see Ref.
\cite{GUP-THR,GUP-THR2}. Thermodynamics of relativistic systems are
studied in the context of DSR theories \cite{DSR-THR}. Moreover, it
is well-known that black holes obey the thermodynamical laws and
emit thermal emission (the so-called Hawking radiation) similar to
the black body radiation \cite{BH}. In the absence of a full quantum
theory of gravity, approaches to quantum gravity proposal such as
string theory and loop quantum gravity reveal some thermodynamical
aspects of the black hole physics \cite{BH-ST,BH-LQG}.
Interestingly, the candidates for flat limit of quantum gravity are
capable to reproduce some important thermodynamical results that are
common between these alternative approaches. For the thermodynamics
of the black holes in noncommutative spaces see Ref. \cite{BH-NC}.
The effects of the generalized uncertainty principle on black holes
thermodynamics are studied in Ref. \cite{BH-GUP}. Therefore, black
holes may be the most interesting thermodynamical systems that is
expected to be considered in the framework of quantum gravitational
statistical mechanics \cite{C-Stat}. In this streamline, we study
thermodynamical properties of a system composed of photon gas in
noncommutative spacetime, as a semiclassical flat limit of quantum
gravity. This study has the potential to open new windows on the
statistical origin of black hole thermodynamics. The results show
that the thermodynamical properties of photon gas will be saturated
at high temperature in noncommutative spacetime. We also find an
interesting analogy between black hole entropy and entropy of
saturated photon gas.

The structure of the paper is as follows: In section \Rmnum{2}, we
present the stable (noncommutative) algebra of relativistic quantum
mechanics and we consider its non-relativistic subalgebra in order
to find the deformed density of states in this setup. In section
\Rmnum{3}, we obtain the modified partition function by means of the
deformed density of states. Using the partition function, we study
thermodynamical properties of the photon gas in noncommutative
spacetime in section \Rmnum{4}. Section \Rmnum{5} is devoted to the
summary and conclusions.

\section{Compact Momentum Space and Deformed Density of States}
In the context of stability theory, special theory of relativity is
understood as a transition from the unstable Lie algebra of Galilean
group, the kinematical group of non-relativistic quantum mechanics,
to the stable algebra of Lorentzian group. Moreover, passage from
classical to quantum mechanics is also appeared to be a transition
from unstable Poisson algebra to the stable Heisenberg one. It was
then pointed out that the fundamental theories of nature may also
seem to be stable in the sense that they do not change in a
qualitative manner under a small change of the associated parameters
\cite{Stable-Th,Stability}. However, when Lorentz and Heisenberg
stable algebras are put together in relativistic quantum mechanics
framework, the resultant algebra turns out to be unstable. It is
then natural to find a stable algebra for the relativistic quantum
mechanics through consideration of structural stability. In this
regard, the minimal candidate for such a stable algebra of
relativistic quantum mechanics is found as \cite{Mendes}
\begin{align}\label{Algebra}
&[J_{\mu\nu},J_{\rho\sigma}]=i(J_{\mu\sigma}\eta_{\nu\rho}+J_{\nu\rho}\eta_{\mu\sigma}
-J_{\nu\sigma}\eta_{\mu\rho}-J_{\mu\rho}\eta_{\nu\sigma}),\nonumber\\
&{[J_{\mu\nu},\,p_\rho\,]}\,=i(p_{\mu}\eta_{\nu\rho}-p_{\nu}\eta_{\mu\rho}),\\
&{[J_{\mu\nu},\,x_\rho\,]}\,=i(x_\mu\eta_{\nu\rho}-x_\nu\eta_{\mu\rho}),\nonumber\\
&{[x_\mu,x_\nu]}=-i\epsilon\,\frac{J_{\mu\nu}}{\kappa^2},\hspace{.5cm}{[p_\mu,x_\nu]}=
i\eta_{\mu\nu}{\mathcal I},\hspace{.5cm}{[p_\mu,p_\nu]}=0,\nonumber\\
&{[x_\mu,\,{\mathcal I}\,]}\,=\,i\epsilon\,\frac{p_\mu}{\kappa^2},\hspace{.9cm}{[p_\mu,
\,{\mathcal I}\,]}\,=0,\hspace{1cm}{[J_{\mu\nu},\,{\mathcal I}\,]}=0,\nonumber
\end{align}
where $\eta_{\mu\nu}=\mbox{diag}(+1,-1,-1,-1)$ is the Minkowski
metric, $\epsilon=\pm1$, and ${\mathcal I}$ is the nontrivial
operator that replaces the trivial center of the standard Heisenberg
algebra \cite{unit}. Clearly, the spacetime coordinates $x_\mu$
become non-commuting operators and the deformation parameter
$\kappa$ with dimension of inverse of length naturally arises in
this setup which could be identified with a minimal length scale.
The deformed algebra (\ref{Algebra}) is Lorentz-invariant through
the preservation of the commutation relation for generators
$J_{\mu\nu}$ and the vector nature of $(x_\mu,p_\mu)$.

It is also useful to compare the stable algebra (\ref{Algebra}) with
the noncommutative spacetime algebra that is proposed by Snyder as
\cite{Snyder},
\begin{align}\label{Snyder-Algebra}
&[J_{\mu\nu},J_{\rho\sigma}]=i(J_{\mu\sigma}\eta_{\nu\rho}+J_{\nu\rho}\eta_{\mu\sigma}
-J_{\nu\sigma}\eta_{\mu\rho}-J_{\mu\rho}\eta_{\nu\sigma}),\nonumber\\
&{[J_{\mu\nu},\,p_\rho\,]}\,=i(p_{\mu}\eta_{\nu\rho}-p_{\nu}\eta_{\mu\rho}),\\
&{[J_{\mu\nu},\,x_\rho\,]}\,=i(x_\mu\eta_{\nu\rho}-x_\nu\eta_{\mu\rho}),\nonumber\\
&{[x_\mu,x_\nu]}=-i\epsilon\,\frac{J_{\mu\nu}}{\kappa^2},\hspace{.5cm}{[p_\mu,p_\nu]}=0,
\nonumber\\&{[p_\mu,x_\nu]}=i\Big(\eta_{\mu\nu}-\epsilon\,\frac{p_{\mu}p_\nu}{\kappa^2}\Big).
\end{align}
The commutator for generators $J_{\mu\nu}$ are the same as the
algebra (\ref{Algebra}), and $(x_\mu, p_\mu)$ transform like a
four-vector which shows that the Lorentz symmetry is also preserved
for the Snyder algebra (\ref{Snyder-Algebra}). In comparison with
the stable algebra (\ref{Algebra}), there is no nontrivial center in
this setup. Interestingly, both of the stable (\ref{Algebra}) and
Snyder (\ref{Snyder-Algebra}) algebras could be obtained from a
ten-dimensional phase space associated to a constrained relativistic
particle in five dimensions in the context of DSR theories. Fixing
the constraint, the stable algebra (\ref{Algebra}) and Snyder
algebra (\ref{Snyder-Algebra}) could be realized from the symplectic
reduction process through the choice of different basis for the
reduced eight-dimensional phase space with curved four-momentum
space (such as the de Sitter or anti-de Sitter geometries)
\cite{Girelli}. In this context, the nontrivial center ${\mathcal
I}$ for the stable algebra (\ref{Algebra}) would be identified with
a five-dimensional coordinate which shows that the algebra
(\ref{Algebra}) is not a closed algebra in four dimensions and it
then does not allow a straightforward four-dimensional
interpretation. The Snyder algebra (\ref{Snyder-Algebra}), however,
is a closed algebra in a four-dimensional spacetime and it could be
considered as the phase space of a four-dimensional spacetime
without any attribution to an extra dimension. There is no clear
reason to prefer stable algebra (\ref{Algebra}) over Snyder algebra
(\ref{Snyder-Algebra}) in DSR framework. But, it seems that the
stable algebra (\ref{Algebra}) is more satisfactory since it could
take into account a universal minimal length in a Lorentz-invariant
manner as well as Snyder algebra (\ref{Snyder-Algebra}) and,
moreover, it is also a stable algebra from the mathematical point of
view. However, when one studies the Heisenberg subalgebra in the
non-relativistic limit to obtain the density of states, depending on
the choice of spatial momenta, as we will show, both of them could
suggest the same modification to the six-dimensional phase space
volume.

A full representation of the stable algebra (\ref{Algebra}) by
differential operators in a five-dimensional manifold with
commutative coordinates ${\xi_A}$ and flat metric
$\eta_{AB}=\mbox{diag}( +1,-1,-1,-1,\epsilon)$ is given by
\cite{mendes2}
\begin{align}\label{representation}
&x_\mu=\frac{i}{\kappa}\Big(\xi_\mu\frac{\partial}{\partial\xi^4}-
\epsilon\xi^4\frac{\partial}{\partial\xi^\mu}\Big),\hspace{1cm}p_\mu
=i\frac{\partial}{\partial\xi^\mu},\\
&J_{\mu\nu}=i\Big(\xi_\mu\frac{\partial}{\partial\xi^\nu}-
\epsilon\xi_\nu\frac{\partial}{\partial\xi^\mu}\Big),\hspace{1cm}
{\mathcal I}=\frac{i}{\kappa}\frac{\partial}{\partial\xi^4}.\nonumber
\end{align}
To obtain the deformed density of states, we restrict ourselves to a
subalgebra containing $\{x^i,p^i,{\mathcal I}\}$ with $i=1,2,3$,
that replaces the standard Heisenberg algebra in this setup. Also we
fix $\epsilon=-1$ since, as we will see, this choice leads to the
compactification of the spatial momenta space and a universal
spatial maximal momentum arises for this case. For a fixed $i$, in
$x$-basis, the representation of this subalgebra is given by
\begin{eqnarray}\label{sub-rep}
x^i=x,\hspace{.7cm}p^i=\kappa\sin\Big(\frac{1}{i\kappa}\frac{d}{
dx}\Big),\hspace{.7cm}{\mathcal I}=\cos\Big(\frac{1}{i\kappa}
\frac{d}{dx}\Big).\hspace{.3cm}
\end{eqnarray}
The states $|p\rangle=e^{ikx}$ would be eigenvectors of $p^i$ with eigenvalues
\begin{equation}\label{pk}
p(k)=\kappa\sin(k/\kappa)\,.
\end{equation}
Considering vanishing boundary condition on a box for $|p\rangle$,
gives $k_n=\frac{\pi}{L}\,n$, with $n\in{\mathbb Z}$. Using the fact
that $dn=\frac{dn}{dp}dp$, from the relation (\ref{pk}), to first
order of approximation the deformed density of states in this setup
will be (see Ref. \cite{Mendes-DOS} for more details),
\begin{equation}\label{DOS}
g(p)dp=\frac{V}{2\pi^2}\frac{p^2dp}{\sqrt{1-(p/\kappa)^2}}\,,
\end{equation}
which is not exact since it is obtained in the spirit of the
subalgebra (\ref{sub-rep}). An important result, which is also clear
from the relation (\ref{pk}), is that there is an spatial maximal
momentum as $p<\kappa$ in this setup. Also, the deformed density of
states (\ref{DOS}) shows that the phase space volume expands at high
momentum regime. Thus, the main result of this section is that: the
stable algebra of relativistic quantum mechanics (\ref{Algebra})
suggests the deformation to the density of states as the relation
(\ref{DOS}). The standard state density could be recovered in the
limit of $\kappa\rightarrow\,\infty$.

The deformed density of states (\ref{DOS}) is also suggested by
three-dimensional DSR as a flat limit for quantum gravity
\cite{Girelli2}. Although the stable (\ref{Algebra}) and Snyder
(\ref{Snyder-Algebra}) algebras are different in four-dimensional
spacetime, the deformed phase space volume which underlies the
density of states (\ref{DOS}) is also obtained from the
non-relativistic (subalgebra) of the Snyder algebra \cite{Mignemi2}.
The reason for this coincidence will become clear when one pays
attention to the topology of the both of four-momentum and spatial
momentum spaces. Indeed, the topology of the four-momentum space of
the Minkowski spacetime with standard non-deformed Poincar\'{e}
algebra is ${\mathbf R}^4$ (flat momentum space) and there is not a
universal maximal spatial momentum or maximum energy for the system
under consideration. But, it is well-known that the topology of the
four-momentum space of the DSR theories is de Sitter geometry with
${\mathbf R}\times{\mathbf S}^3$ \cite{AdsSP}. Identifying ${\mathbf
R}$ with energy and ${\mathbf S}^3$ with spatial momenta, a
universal maximal momentum (corresponds to a minimal
observer-independent length) naturally arises which is completely
determined by the radius of three-sphere ${\mathbf S}^3$ or
equivalently with curvature of de Sitter four-momentum space. Since
both of the stable algebra (\ref{Algebra}) and Snyder algebra
(\ref{Snyder-Algebra}) could be realized from the de Sitter (curved)
four-momentum space in the context of DSR, a universal maximal
momentum naturally arises in both of these setups through the
identification of ${\mathbf S}^3$ topology with the space of spatial
momenta. Also, it was already pointed out that the phase spaces with
compact topology for the momentum part are naturally
ultraviolet-regularized \cite{LT}. This is because a Liouville
volume associated to a compact phase space is finite and the
corresponding Hilbert space is finite-dimensional \cite{Rovelli}.
Moreover, a measure very similar to (\ref{DOS}) is suggested by the
classical limit \cite{CPR} of polymer quantum mechanics \cite{QPR}.
However, considering the polymerized systems within the boundaries
of the standard methods in statistical mechanics is in some sense
debatable \cite{PLR-OS}.

Thus, the density of states (\ref{DOS}) underlies the ${\mathbf
S}^3$ compact topology and consequently a universal maximal momentum
$p<\kappa$, as an ultraviolet cutoff for the system under
consideration, naturally arises in this setup. This maximal momentum
is completely determined by the radius of three-sphere or
equivalently by the curvature of de Sitter four-momentum space in
which three-sphere is embedded. As we will show in the next
sections, existence of a maximal momentum leads to the saturation of
thermodynamical properties of the statistical systems at the high
temperature regime.

\section{Partition Function}
In this section, we ob ain the modified partition function by means
of the deformed density of states (\ref{DOS}) in order to study
thermodynamics of photon gas in noncommutative spacetime.

The number of accessible microstates for a physical system
determines the associated thermodynamical properties in the context
of statistical mechanics. The microstates, however, is determined
only by quantum mechanics and there is no classical statistics in
essence. More precisely, for a system at temperature $T$, the
quantum partition function for a single-particle state is defined as
\begin{equation}\label{QPF}
{\mathcal Z}_1=\sum_{\varepsilon_i}e^{-{\varepsilon_i}/T}\,,
\end{equation}
where $\{\varepsilon_i\}$ are the single-particle energy states
which are the solution of Hamiltonian eigenvalue problem. From the
single-particle partition function one could find the total
partition function and all the thermodynamical properties of the
system under consideration then could be easily obtained from the
total partition function. Solving the Hamiltonian eigenvalue problem
in noncommutative spacetime, however, is not an easy task at all due
to the complicated form of the representations of the operators in
this setup (see for instance Ref. \cite{Mendes-other}). Even if one
solves the Hamiltonian eigenvalue problem, performing the summation
over the microstates' energies in relation (\ref{QPF}) to obtain an
analytic expression for the associated quantum partition function
may be difficult. Nevertheless, one could approximate the summation
over the energies by the integral over the phase space volume by
means of the density of states. The advantage of this method is
that, it is not need to solve the Hamiltonian eigenvalue problem in
noncommutative setup. Indeed, one could work with the Hamiltonian
function together with the deformed density of states (\ref{DOS}) on
the associated noncommutative phase space. However, one should note
that this semiclassical approximation will be coincided with full
quantum consideration at high temperature regime and the full
quantum considerations then preserve their importance for the low
temperature phenomenons such as the Bose-Einstein condensation.
Thus, all the noncommutativity effects are summarized in the
deformed density of states (\ref{DOS}) and one could replace the
summation over the microstates' energies as
\begin{eqnarray}\label{Approx}
\sum_{\varepsilon}\rightarrow\frac{V}{(2\pi)^3}\int_{|p|\leq{\kappa}}
\frac{d^3p}{\sqrt{1-\frac{p^2}{\kappa^2}}}\,,
\end{eqnarray}
where $V$ is the spatial volume. Using the semiclassical
approximation (\ref{Approx}) in the relation (\ref{QPF}), the
canonical partition function will be
\begin{eqnarray}\label{PF-D}
{\mathcal Z}_1(T,V)=\frac{1}{(2\pi)^3}\int\,d^3x\int_{|p|\leq{\kappa}}
\frac{e^{-H(x,p)/T}\,d^3p}{\sqrt{1-(p/\kappa)^2}}.
\end{eqnarray}
For a photon gas confined in a volume $V$ at temperature $T$ with
Hamiltonian function $H(p)=p$, the associated modified partition
function for the single-particle state could be easily deduced from
the relation (\ref{PF-D}) as
\begin{eqnarray}\label{PF-D2}
{\mathcal Z}_1(T,V)=\frac{V}{(2\pi)^3}\int_{0}^{\kappa}\frac{e^{-p/T}\,p^2dp}{
\sqrt{1-(p/\kappa)^2}}=\frac{V}{(2\pi)^3}\,f(T),\hspace{.5cm}
\end{eqnarray}
where we have defined
\begin{eqnarray}
f(T)=-\frac{\kappa^4}{3}T+\frac{\pi\kappa^2}{2T}\bigg\{\Big[
I_1(\kappa{T})+\kappa{T}\,I_2(\kappa{T})\Big]-\Big[{\bf L}_1(
\kappa{T})+\kappa{T}\,{\bf L}_2(\kappa{T})\Big]\bigg\},\hspace{.5cm}
\end{eqnarray}
with $I_n$ and ${\bf L}_n$ are the modified Bessel functions and
modified Struve functions of rank $n$ respectively. In the limit of
low temperature, the noncommutativity effects will be removed and we
find $f(T)=\,8\pi\,T^3$ and the relation (\ref{PF-D2}) reduces to
the standard non-deformed partition function for the photon gas
${\mathcal Z}_1(T\rightarrow{0},V)\approx8\pi{V}(T/h)^3$. The total
partition function for a non-localized system such as photon gas,
with which we are interested in this paper, can be obtained from the
standard definition
\begin{eqnarray}\label{TPF-IG-D}
{\mathcal Z}_{N}(T,V)=\frac{1}{N!}\big[{\mathcal Z}_1
(T,V)\big]^N\,,
\end{eqnarray}
where $N$ is the number of photons and the Gibb's factor is also
considered. Substituting the single-particle state partition
function (\ref{PF-D2}) into the definition (\ref{TPF-IG-D}) gives
the total partition function for the photon gas as
\begin{eqnarray}\label{TPF-IG}
{\mathcal Z}_{N}(T,V)=\frac{V^N}{(2\pi)^{3N}}\,\frac{f(T)^N}{N!}\,.
\end{eqnarray}

Having the total partition function (\ref{TPF-IG}) in hand, we are
able to study thermodynamics of the photon gas in noncommutative
spacetime. Considering the maximal spatial momentum $\kappa$ to be
of the order of the Planck momentum (energy) as
$\kappa\sim\,p_{_{\rm Pl}}$, it is natural to expect that the
deviation from the standard thermodynamical results for a photon gas
emerges at the high temperature regime (about the Planck
temperature) where the noncommutative effects play significant role.

\section{Thermodynamics of Photon Gas}
First of all, we obtain the Helmholtz free energy $F$ which can be
obtained from the standard definition as
\begin{eqnarray}\label{Helmholtz}
F=-T\ln\big[{\mathcal Z}_{N}(T,V)\big]=-NT\left[1+\ln\Big(\frac{Vf}{8\pi^3N}\Big)\right],
\hspace{.5cm}
\end{eqnarray}
in which we have used the Stirling's formula
$\ln(N!)\approx{N}\ln{N}-N$ for large $N$. The thermal pressure of
the photon gas will be
\begin{eqnarray}\label{pressure}
P=-\bigg(\frac{\partial F}{\partial V}\bigg)_{T,N}=\frac{NT}{V}\,.
\end{eqnarray}
Therefore, the familiar form of the equation of state for the ideal gasses
\begin{equation}\label{EOS}
PV=NT,
\end{equation}
is also preserved in this setup. It is the well-known result that
the from of equation of state (\ref{EOS}) is held for both of the
relativistic and non-relativistic ideal gases in standard
Maxwell-Boltzmann statistics. It seems that this form also remains
unchanged for both of the relativistic and non-relativistic ideal
gases in the presence of an ultraviolet cutoff (see for instance
Refs. \cite{GUP-THR2,Gorji} for the case of generalized uncertainty
principle setup and noncommutative spaces).
\subsection{Entropy Bound}
\begin{figure}
\flushleft\leftskip+15em{\includegraphics[width=2.5in]{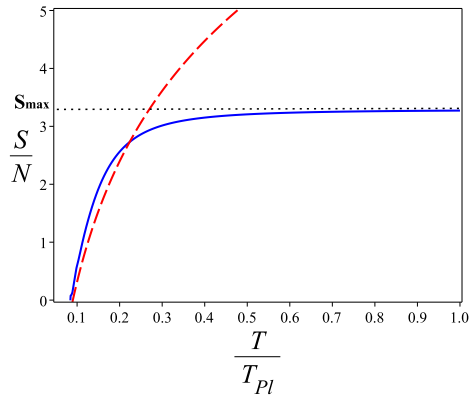}}
\hspace{3cm}\caption{\label{fig:1} Entropy versus the temperature.
The solid line represents the entropy of the photon gas in
noncommutative spacetime with maximal momentum for the spatial
momenta and the dashed line corresponds to the non-deformed case. At
the high temperature regime, the entropy increases with a slower
rate in noncommutative spacetime with respect to the corresponding
non-deformed case. At high temperature, the entropy approaches to a
maximum value (\ref{max-entropy}) which is originated from the
existence of an ultraviolet cutoff (maximal momentum corresponds to
a minimal length) in noncommutative framework. The figure is plotted
for $T_{_{\rm Pl} }=1$, where $T_{_{\rm P}}$ is the Planck
temperature.}
\end{figure}

The entropy could be obtained from the Helmholtz energy (\ref{Helmholtz}) by definition as
\begin{eqnarray}\label{entropy}
S=-\bigg(\frac{\partial F}{\partial T}\bigg)_{N,V}=N\Big(1+
\ln(V/8\pi^3N)+(T\ln{f})'\Big),\hspace{.7cm}
\end{eqnarray}
where a prime denotes the derivative with respect to the
temperature. The behavior of the entropy versus the temperature is
plotted in figure \ref{fig:1}. It is clear from figure \ref{fig:1}
that the entropy increases with a slower rate in noncommutative case
(the solid line) than the standard one (the dashed line). Unlike the
entropy of the standard photon gas, the entropy approaches to a
maximum value at very high temperature regime in the noncommutative
setup. The maximum entropy bound for the photon gas is
\begin{eqnarray}\label{max-entropy}
S\leq{S}_{_{\rm max}}=C_1(N)+N\ln\bigg(\frac{V}{l_{_{\rm Pl}}^3}\bigg)\,,
\end{eqnarray}

where $C_1(N)=N+N\ln\Big(\frac{\kappa_0^3}{8\pi{N}}\Big)$ and also
we have substituted $\kappa= \kappa_0\,p_{_{\rm
Pl}}=\kappa_0/l_{_{\rm Pl}}$ with $\kappa_0$ is the dimensionless
parameter and $l_{_{\rm Pl}}$ is the Planck length. The
dimensionless parameter $\kappa_0\sim{\mathcal O}(1)$ determines the
boundary at which the noncommutative effects will become important
and it should be fixed only by the experiments (see Ref.
\cite{QGExperiment} in which some upper bounds for this
dimensionless parameter are obtained in different contexts). The
entropy bound (\ref{max-entropy}) shows that the photon gas is
saturated at high temperature regime in noncommutative spacetime and
the entropy could not increase by increasing the temperature in this
regime. More precisely, the entropy bound (\ref{max-entropy}) for
the photon gas originates from the existence of an ultraviolet
cutoff in this setup which also, as we shall see, leads to an upper
bound for the internal energy.

Appearance of the factor $\big(V/l_{_{\rm Pl}}^3\big)$ in the
relation (\ref{max-entropy}) for the maximum entropy bound, signals
the discreteness of the space with respect to the Planck length at
high temperature limit in this setup. This result, in some sense, is
similar to the case of the entropy of the black holes. The
Bekenstein-Hawking entropy for black holes is given by the relation
$S_{_{BH}}=\big(A/4l_{_{\rm Pl}}^2\big)$, where $A$ is the horizon
area of black hole \cite{BH}. Then, the number of microstates is
precisely determined by the horizon area and the fundamental area
$l_{_{\rm Pl}}^2 $. In other words, each bit of information is
proportional to the fundamental (Planck) area $l_{_{\rm Pl}}^2$. In
the case of the photon gas in noncommutative setup, the entropy
bound (\ref{max-entropy}) shows that the number of microstates at
high temperature regime is precisely determined by the factor
$\big(V/l_{_{\rm Pl}}^3\big)$, {\it i.e.}, one bit of information is
proportional to $l_{_{\rm Pl}}^3$. In other words, similar to the
black holes, the entropy at high energy regime (where quantum
gravitational effects play the central role) precisely is determined
by the accessible spatial volume together with the ultraviolet
cutoff (Planck length) for the systems. Many attempts have been done
to find the microscopic origin of black holes entropy
\cite{BH-LQG,BH-ST,BH-Stat}. Then, in the light of this result, it
may be possible that black holes thermodynamics will emerge from a
saturated state in a full quantum gravitational statistical
mechanics framework. In the absence of such a theory, the
gravitational statistical theories such as the generally covariant
statistical mechanics may open a new window on this issue
\cite{C-Stat}.
\subsection{Internal Energy and Specific Heat}
\begin{figure}
\flushleft\leftskip+15em{\includegraphics[width=2.5in]{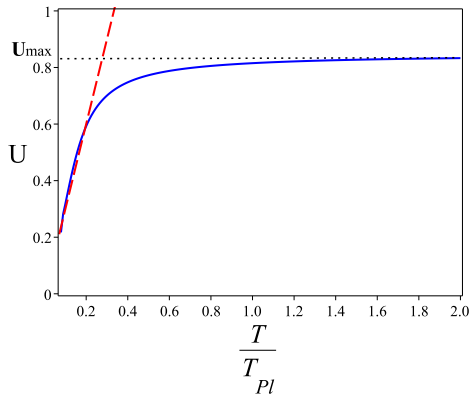}}
\hspace{3cm}\caption{\label{fig:2} Internal energy of the photon gas
versus the temperature. The solid line represents the internal
energy in noncommutative setup and the dashed line corresponds to
the standard case. Clearly, noncommutative effects dominate when the
temperature approaches the Planck temperature and the internal
energy of the photon gas then increases with a much less rate in
this regime. The system will be finally saturated and the internal
energy approaches to its maximum value (\ref{max-energy}) which is
of the order of Planck energy. This feature also originates from the
existence of an ultraviolet cutoff in this setup.}
\end{figure}

The internal energy of the photon gas will be
\begin{eqnarray}\label{energy}
U=-T^2\bigg[\frac{\partial}{\partial T}\bigg(\frac{F}{
T}\bigg)\bigg]_{N,V}=NT^2\,\frac{f'}{f}.
\end{eqnarray}
Substituting $f\sim\,T^3$ for the low temperature regime,
immediately gives the standard result $U=3NT$ for the photon gas.
The internal energy versus the temperature is plotted in figure
\ref{fig:2}. It is clear from the figure \ref{fig:2} that, at high
temperature regime, the internal energy of the photon gas in
noncommutative framework (the solid line) increases with a smaller
rate with respect to the non-deformed case (the dashed line).
Interestingly, there is an upper bound, of the order of the Planck
energy, for the internal energy of the photon gas in noncommutative
spacetime as follows
\begin{eqnarray}\label{max-energy}
U\leq\,U_{_{\rm max}}=\Big(\frac{8}{9\pi\kappa_0}\Big)
E_{_{\rm Pl}}\,.
\end{eqnarray}
This result also originates from the fact that there is an
ultraviolet cutoff $|p|\leq\kappa$ for the system under
consideration in noncommutative spacetime.

\begin{figure}
\flushleft\leftskip+15em{\includegraphics[width=2.5in]{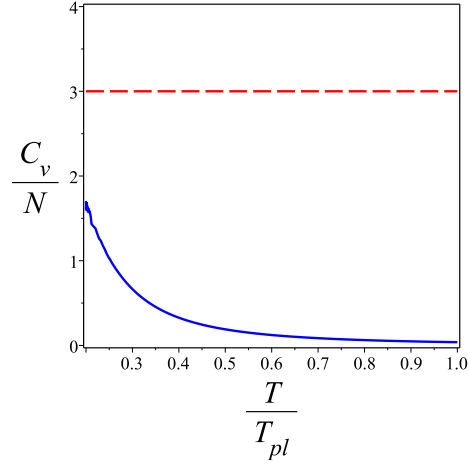}}
\hspace{3cm}\caption{\label{fig:3} Specific heat versus the
temperature. The solid line represents the heat capacity of the
photon gas in noncommutative space and the dashed line corresponds
to the non-deformed case. While the specific heat is constant (for a
fixed $N$) for the standard photon gas, it becomes
temperature-dependent at high temperature regime in noncommutative
spacetime. Indeed, the photon gas saturates at the high energy
regime and then the specific heat tends to zero in this regime. In
other words, the system cannot access a higher energy scale by
increasing the temperature.}
\end{figure}

The specific heat then will be
\begin{eqnarray}\label{SH}
C_{_V}=\bigg(\frac{\partial U}{\partial T}\bigg)_{V}=NT^2\bigg(
\frac{f''}{f}-\frac{f'^2}{f^2}+\frac{2f'}{Tf}\bigg)\,.
\end{eqnarray}
The specific heat versus the temperature is plotted in figure
\ref{fig:3}. Unlike the standard photon gas with constant specific
heat $C_{_V}=3N$ (for a fixed $N$), the specific heat in
noncommutative framework becomes temperature-dependent at high
energy regime (see figure \ref{fig:3}). As it is clear from the
figure \ref{fig:3}, $C_{_V}(T\rightarrow\infty)=0$ which shows that
the photon gas saturates at high temperature limit (around the
Planck temperature), when the noncommutative effects become
significant and the system cannot access a higher energy scale by
increasing the temperature.

\section{Summary and Conclusions}

Existence of a minimal length, preferably of the order of Planck
length, is suggested by quantum gravity candidates such as string
theory and loop quantum gravity. Although a full quantum theory of
gravity is not formulated yet, it is widely believed that a
non-gravitational theory that admits a minimal length scale would be
emerged at the flat limit of quantum gravity. Evidently, such a
theory could be achieved through the deformation of the algebraic
structure of the standard relativistic quantum mechanics in such a
way that spacetime coordinates become non-commutating operators.
Recently, in the context of doubly special relativity theories, it
was shown that this issue could be also realized from a curved
four-momentum space with constant curvature such as the de Sitter
geometry with topology ${\mathbf R}\times{\mathbf S}^3$. Identifying
${\mathbf R}$ with the space of energy and ${\mathbf S}^3$ with the
space of spatial momenta, a universal maximal momentum (corresponds
to a minimal observer-independent length scale) naturally arises
which is completely determines by the radius of three-sphere or
equivalently with curvature of the de Sitter four-momentum space.
The deformation to the dispersion relation and density of states are
the direct consequence of these setups. Since the number of
microstates is determined by the density of states, the significant
effects on the thermodynamical properties of the physical systems
will be arisen in these setups. In this paper, after obtaining the
modified partition function by means of the deformed density of
states in spacetime with stable noncommutative algebra, we have
studied the thermodynamical properties of the photon gas in this
setup. The results show that the entropy of the photon gas increases
with a smaller rate at the high temperature regime in noncommutative
setup in comparison with the standard non-deformed case. Also, the
entropy approaches to a maximum entropy bound around the Planck
temperature. The number of accessible microstates associated to the
resultant entropy bound is totally determined by the factor
$\big(V/l_{_{\rm Pl}}^3\big)$ which is qualitatively very similar to
the black holes' Bekenstein-Hawking entropy $S_{BH}=\big(A/l_{_{\rm
Pl}}^2\big)$ in which the number of accessible microstates for a
black hole is precisely determined by the factor $\big(A/l_{_{\rm
Pl}}^2 \big)$, where $A$ is the black hole horizon area. In other
words, similar to the case of black holes, the entropy for the
photon gas in high temperature regime is precisely determined by the
accessible spatial volume and an ultraviolet cutoff in
noncommutative spacetime. This result suggests that thermodynamics
of black holes may be obtained from a saturated state in the
framework of a full quantum gravitational statistical mechanics.
Furthermore, the internal energy of the photon gas also gets a
finite maximum value, of the order of the Planck energy, at very
high temperature regime. The associated specific heat then tends to
zero at high temperature regime since the photon gas is saturated in
this regime and could not access to a higher energy scale by
increasing the temperature. Although we have studied the
thermodynamics of a particular system (the photon gas) in
noncommutative spacetime, having upper bounds for the entropy and
internal energy are generic properties since they originate from the
existence of an ultraviolet cutoff in this setup.


\begin{thebibliography}{1}
\bibitem{unit} We work in unit $\hbar=c=
k_{_B}=1$, where $\hbar$, $c$ , and $k_{_B}$
are the Planck constant, speed of light in
vacuum, and Boltzman respectively.
\bibitem{AdsSP} More precisely, the space
of four-momentum will be maximally symmetric
with constant curvature and the anti-de
Sitter geometry then also be a well-defined
candidate. See Refs. \cite{DOS-Ads,AdS}
for more details.
\end{thebibliography}
\end{document}